\documentclass[pra,aps,showpacs,epsf, twocolumn]{revtex4}

\def\be{\begin{equation}}
\def\ee{\end{equation}}

\usepackage[dvips]{graphicx}
\usepackage{bm}
\usepackage{amsmath}
\begin{document}

\title{Ground state of a tightly bound composite dimer immersed in a Fermi Sea}
\author{Christophe Mora$^{1}$ and Fr\'ed\'eric Chevy$^{2}$}
 \affiliation{$^{1}$Laboratoire Pierre Aigrain, \'Ecole Normale
Sup\'erieure and CNRS,
Universit\'e Paris 7 Diderot; 24 rue Lhomond, 75005 Paris, France}

 \affiliation{$^{2}$Laboratoire Kastler Brossel, CNRS, UPMC, \'Ecole Normale Sup\'erieure, 24 rue Lhomond, 75231 Paris, France}
\date{\today}

\begin{abstract}
In this paper we present a theoretical investigation for the ground state of an impurity immersed in a Fermi sea.
The molecular regime is considered where a two-body bound state between the impurity and one of the fermions is formed.
Both interaction and exchange of the bound fermion take place between the dimer and the Fermi sea.
We develop a formalism based on a two channel model allowing us to expand systematically the ground state energy of this immersed dimer with the scattering length $a$. Working up to order $a^3$, associated to the creation of two particle-hole pairs, reveals the first signature of the composite nature of the bosonic dimer.
Finally, a complementary variational study provides an accurate estimate of the dimer energy even at large scattering length.
\end{abstract}

 \maketitle

\section{Introduction}

As demonstrated by Bardeen-Cooper and Schrieffer, superconductivity arises from the pairing of electrons with opposite spins into Cooper pairs \cite{BCStos1957}. A natural extension of their work to the case where the two spin populations are imbalanced was proposed by Clogston and Chandrasekhar \cite{clogston1962ulc,chandrasekhar1962}: they suggested that when a magnetic field is a applied, the existence of a pairing gap in the electron energy spectrum could prevent spin polarization as long as the Zeeman shift was smaller than the gap, a threshold known as the Clogston-Chandraskhar (CC) limit. It was later suggested that the superfluid state could survive beyond the CC limit, into the form of a non-homogeneous superconducting state known as the FFLO (Fulde Ferrell Larkin Ovchinnikov) state \cite{Fulde1964sss,larkin1965}. However, most superconductors behave like nearly ideal diamagnetic compounds (Meissner Effect), which forbids magnetically induced spin polarization in the bulk, see Ref.~\cite{matsuda07} for a short review. As a consequence, theses theories were investigated experimentally only very recently in a series of works performed at Rice \cite{partridge2006pap} and MIT \cite{zwierlein2006fsi} with ultra-cold Fermi gases trapped in optical potentials. Although some debate on the structure of the normal component persists between the two groups, both observe a shell structure in the density profile, with at center a fully paired region consistent with the CC scenario of robust fermionic superfluidity.

The main discrepancy between the two experiments lies in the polarization of the normal  component: Indeed, while Rice's group observed that the outer rim was composed exclusively of majority atoms, the normal component obtained at MIT was only partially polarized and contained also particles of the minority spin species. Recent theoretical work demonstrated that this latter observation was compatible with the homogeneous phase diagram of a strongly interacting Fermi gas if one assumes the validity of the Local Density Approximation \cite{pilati2008psi}. In particular, it was shown that some of the most salient features  could be understood fairly accurately from the study of the simpler problem of an impurity immersed in a Fermi sea \cite{bulgac07ztt,chevy2006upa}. At unitarity (scattering length $a=\infty$), it was demonstrated that the impurity could be  described as a quasi-particle dressed by particle hole excitations of the background Fermi sea \cite{chevy2006upa,combescot2007nsh,combescot2008nsh,lobo2006nsp}. Comparison between simplified variational models and Monte-Carlo simulations have in addition shown that even in this regime of strong correlation, a single particle-hole excitation was sufficient to capture quantitatively the properties of this so-called Fermi polaron \cite{lobo2006nsp,prokof'ev08fpb}. However, the Fermi-polaron picture is valid only for $a$ negative, and close to unitarity $1/k_F|a|\gg 1$. In the $1/k_Fa\lesssim 1$ regime, Monte-Carlo simulations have revealed that the Fermi-polaron was not the ground state of the system anymore \cite{prokof'ev08fpb}. Indeed, for $k_Fa < 1.11$, the ground-state is now described by a bosonic dimer interacting with the background Fermi sea where the atom-dimer scattering length~\cite{petrov2003} $a_{dm}\simeq 1.1786 \,a$ was first calculated by Skorniakov and Ter Martirosian \cite{Skorniakov1957}.

In this paper, we use a combination of perturbative expansion and variational calculation to draw a simple and intuitive picture of the molecular sector of the impurity problem and extend diagrammatic calculations presented in \cite{pieri06tfd,combescot09atd}: we show that for $a>0$, the molecular impurity shares several features with the Fermi-polaron describing the impurity for $a<0$. In particular, we show that the creation of a single particle-hole pair in the Fermi sea provides an accurate quantitative description of the system. The paper is organized as follows: Sec.~\ref{Main} presents the main results of the paper. More technical aspects of this work are then relegated to following sections, Sec.~\ref{Perturbative} for the perturbative calculation and Sec.~\ref{Variational} for the variational approach. Sec.~\ref{conclusion} concludes.

\section{Main results}
\label{Main}

\begin{figure}
\centerline{\includegraphics[width=\columnwidth]{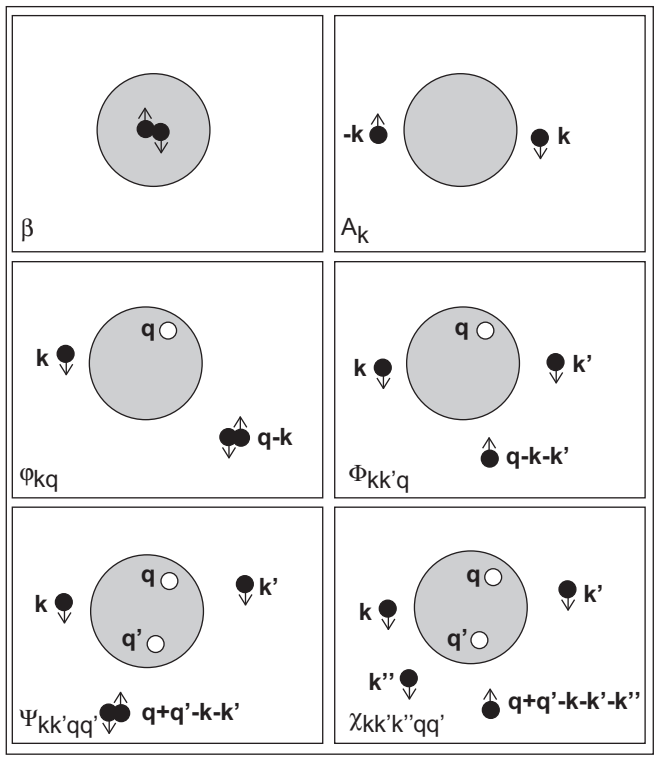}}
\caption{Our variational scheme takes into account only a restricted number of particle-hole excitations. First row: no particle hole excitation. Second row: one particle-hole excitation of momenta $q<k_F$ and $k>k_F$. Third row: two particle-hole excitations of momenta $q,q'<k_F$ and $k,k'>k_F$.}
\label{Fig0}
\end{figure}

Let us consider an ensemble of spin $F=1/2$ fermions (for instance $^6$Li in its hyperfine ground state), where all particles but one are polarized in $m_F=-1/2$. At low temperature, short range interatomic interactions between identical fermions are suppressed by Pauli exclusion principle, and the system can therefore be described as a single impurity interacting with a Fermi sea of non-interacting fermions. We note $a$ the scattering length between particles of opposite spins. The case $a<0$ and the vicinity
of the Feshbach resonance ($a \to \infty$) have been described in previous works on the Fermi-polaron \cite{chevy2006upa,combescot2007nsh,combescot2008nsh,lobo2006nsp}, and here we concentrate on the molecular sector corresponding to $a$ small and positive. In this particular regime the picture is relatively simple, since in the absence of a surrounding Fermi sea, the two body potential possesses a deeply bound state of size $\sim a$ and energy $E=-\hbar^2/ma^2$. When one adds a spin $m_F=+1/2$ atom to the Fermi sea, it will pair up with a majority atom to form a deeply bound dimer. In the regime $k_F a\ll 1$, the size of this molecule is much smaller than the inter-particle distance, and we can describe the dimer as a point like boson of mass $2m$. In this picture, the energy shift associated with the addition of the impurity is then at leading order in $k_F a$

$$\Delta E=-\frac{\hbar^2}{m a^2}-E_F+...,$$

\noindent where the second term corresponds to the removal of one majority fermion from the Fermi sea to form the bosonic dimer. The next order in the $k_Fa$ expansion comes from the interactions between the bosonic dimer and the surrounding Fermi sea. Indeed, if the point-like boson picture is correct, then one should expect a mean-field energy shift $\Delta E=g_{\rm ad} n$, where $n$ is the density of majority atoms and $g_{\rm ad}$ is the atom-dimer coupling constant given by $g_{\rm ad}=3\pi\hbar^2 a_{\rm ad}/m$ associated with the atom-dimer scattering length $a_{\rm ad}\sim 1.1786 a$ \cite{petrov2003,Skorniakov1957}. Although this scenario has been confirmed by Monte-Carlo simulations \cite{prokof'ev08fpb}, the analytic calculation of the mean field shift is not trivial for two main reasons. First, the atom-dimer scattering length is obtained by solving the three body-problem, but here antisymmetrization of the global wave-function correlates automatically the majority atom bound in the dimer with the surrounding Fermi-sea, and turns the calculation from three to many-body. Second, mean field contributions are usually obtained using a perturbation expansion based on the existence of a small parameter (here $k_F a$), and in our situation, this scheme is due to fail. Indeed, in absence of interactions ($a=0$), the system is just an ensemble of non interacting fermions, that cannot form any dimer, which is contradictory with the intuitive picture we drew above. To circumvent this latter point, we decided to work in the two-channel picture \cite{jona2008,gogolin2008asb}, where the short-range two-body bound state responsible for the Feshbach resonance is explicitly included in the model as a bosonic degree of freedom, of mass $2m$ and bare binding energy $E_0$.  In second quantized form, the Hamiltonian reads
\begin{equation}\label{hamil}
\begin{split}
H =& \sum_{\bm k,\sigma=\uparrow,\downarrow} \epsilon_k a_{\bm k,\sigma}^\dagger
a_{\bm k,\sigma}^{} + \sum_{\bm K} (E_0+ \epsilon_K /2 )
b_{\bm K}^\dagger b_{\bm K}^{} \\
+&
\sum_{{\bm k},{\bm K}} \frac{\Lambda_{k}}{\sqrt{V}} \left( b^\dagger_{\bm K} a^{}_{{\bm k}+{\bm K}/2,\uparrow}
a_{-{\bm k}+{\bm K}/2,\downarrow}^{} + {\rm h.c.} \right),
\end{split}
\end{equation}

\noindent where $V$ is a quantization volume and $\Lambda_k$ is the matrix element coupling fermionic and bosonic degrees of freedom. If $r_b$ is the typical size of the bare molecule, then the width of $\Lambda_k$ is $\sim 1/r_b$.
The fermionic operator
$a_{{\bm k},\sigma}$  describes a momentum-${\bm k}$ atom
with spin  (hyperfine state) $\sigma$. Atoms in the two
different spin states,  $\sigma=\uparrow,
\downarrow$, are coupled {\it via} a molecular closed
channel state described by the boson operator $b_{\bm K}$
with bare energy at rest $E_0$ and momentum ${\bm K}$.
With an even  (and very weak, see below) $k$-dependence
for the coupling strength
$\Lambda_{k}$, the resulting atomic interaction has an s-wave
character. In the universal regime where the properties of the system depend only on the scattering length, no other ingredient is necessary, as long as $E_0$ and $\Lambda_k$ are chosen to reproduce the actual interatomic scattering length. The Pauli principle imposes vanishing s-wave
interaction for fermions in the same spin channel and do not require coupling between same spin particles.

Even in the case where a single spin down particle is immersed in a Fermi sea of spin up particles, this Hamiltonian does not have a complete analytical solution. One needs to rely on approximations to get some insights on the behavior of the system. It can be assumed, in analogy with the Fermi-polaron problem, that the most salient features are obtained by taking into account the excitations of a small number of particle-hole pairs of the background Fermi sea. In the work presented here, we take into account up to two particle-hole pairs, which reveals effects due to the composite nature of the bosonic dimer. The variational state that we have studied thus takes the form (see also Fig. \ref{Fig0})
\begin{equation}\label{improvedansatz4}
\begin{split}
& |\psi\rangle =  \Bigg( \beta  b_{{\bm K}=0}^\dagger + {\sum_{\bm k}}' A_{\bm k}
a^\dagger_{{\bm k},\downarrow} a^\dagger_{-{\bm k},\uparrow} \\[1mm]
&  - {\sum_{{\bm k},{\bm q}}}' \varphi_{{\bm k},{\bm q}}
a^\dagger_{{\bm k},\downarrow} b_{-{\bm k}+{\bm q}}^\dagger a_{{\bm q},\downarrow} \\[2mm]
& - \! \!   {\sum_{{\bm k},{\bm k}',{\bm q}}}'
  \Phi_{{\bm k},{\bm k'},{\bm q}}
a^\dagger_{{\bm k},\downarrow}
a^\dagger_{{\bm k}',\downarrow} a^\dagger_{-{\bm k'}-{\bm k}+{\bm q},\uparrow}
a_{{\bm q},\downarrow} \\[2mm]
& - {\sum_{{\bm k},{\bm k}',{\bm q},{\bm q}'}}' \Psi_{{\bm k},{\bm k}',{\bm q},{\bm q}'}
a^\dagger_{{\bm k},\downarrow}
a^\dagger_{{\bm k}',\downarrow} b^\dagger_{-{\bm k'}-{\bm k}+{\bm q}+{\bm q}'}
a_{{\bm q},\downarrow} a_{{\bm q}',\downarrow} \\[2mm]
& - {\sum_{{\bm k},{\bm k}',{\bm k}^{\prime \prime},{\bm q},{\bm q}'}}'
\chi_{{\bm k},{\bm k}',{\bm k}^{\prime \prime},{\bm q},{\bm q}'}
a^\dagger_{{\bm k},\downarrow}
a^\dagger_{{\bm k}',\downarrow} a^\dagger_{{\bm k}'',\downarrow} \\[2mm]
& a^\dagger_{-{\bm k'}-{\bm k}-{\bm k}^{\prime \prime}+{\bm q}+{\bm q}',\uparrow}
a_{{\bm q},\downarrow} a_{{\bm q}',\downarrow}  \Bigg) |FS\rangle,
\end{split}
\end{equation}
where  $\sum'$ means that the sum over the majority particle (resp. hole) wavevectors $k$ (resp. $q$) is restricted to $k>k_F$ (resp. $q<k_F$),  and $|FS\rangle$ is the noninteracting Fermi sea of majority atoms, in the absence of minority particle or short range molecule. Solving exactly the equations in this restricted subspace
is a numerical difficult task outside the scope of this work. We therefore simplify the problem in two directions: one leads us to a  systematic {\it perturbative}
expansion in $k_F a$, the other one to a {\it variational} prediction for the molecule energy.

We first expand the variational equations in power of $k_Fa$. In fact, as it was stressed in \cite{combescot2008nsh}, an expansion in $k_Fa$ is equivalent to an expansion in the number of excited particle-hole pairs. In particular, the variational space~\eqref{improvedansatz4} is sufficient to derive the exact and  systematic
expansion of the immersed molecule  energy up to order $(k_F a)^3$. To be more precise, extending the variational space by allowing additional electron-hole pairs
would lead to corrections that are higher orders in $k_F a$.

Interestingly, the mathematical structure of the perturbative equations exhibits a separation of
length scales, $k_F a \ll 1$, which can be given a simple physical interpretation. $\varphi_{{\bm k},{\bm q}}$ is seen as the atom-dimer
wavefunction when the Fermi sea is composed of a single majority fermion. It presents a singularity
$\sim 1/k^2$ at small $k$ corresponding to the general large distance decay $\sim 1/r$ for the scattered wave of a short-range potential. The prefactor gives
the renormalized atom-dimer scattering length. Due to this singular behaviour, the wavefunction $\varphi_{{\bm k},{\bm q}}$ is modified separately
at the two length scales $a$ and $\lambda_F=1/k_F$ in the presence of the Fermi sea.
It is first modified at small wavevectors $\sim k_F$, {\it i.e.} close to the singularity, the stronger effect being Pauli blocking which restricts allowed 
wavevectors ${\bm k}$ to lie outside the Fermi sea. The corresponding contributions to the molecule energy are the same as in the point-like boson case,
like the first two corrections in Eq.~\eqref{ener5} for example. They involve large distances $\sim \lambda_F \gg a$ over which the composite nature
of the molecule is not visible. In contrast to that, corrections to $\varphi_{{\bm k},{\bm q}}$ for larger wavevectors $\simeq 1/a$ reveal the composite
structure of the molecule and change the value of $\alpha_3$ in Eq.~\eqref{ener5}.

Using the simplifications permitted by the length-scale separation mentioned above and working out the expansion in $x_F=k_Fa$ to third order yields
 the following expression for the molecule energy~\footnote{Note that the energy origin is chosen so that $|FS\rangle$ has zero energy and this convention differs by $E_F$ from that used in \cite{prokof'ev08fpb}.} (see Eq.~\eqref{ener4})
\begin{equation}\label{ener5}
\begin{split}
E  = - \frac{\hbar^2}{m a^2} & \Bigg( 1 - \frac{x_F^3 \tilde a_{\rm ad}}{2 \pi}
- \frac{2 x_F^4 \tilde a_{\rm ad}^2}{\pi^2} ( \ln 2 - 3/8) \\[2mm]
& + 2 x_F^5 \alpha_3  +...\Bigg),
\end{split}
\end{equation}
where $\tilde a_{\rm ad}=a_{\rm ad}/a\sim 1.1786$ is the ratio between the atom-dimer and atom-atom scattering lengths, and $\alpha_3 \simeq 0.0637$ is a numerical coefficient obtained from the formalism described in Sec.~\ref{Perturbative}. As mentioned above, the first term of the expansion corresponds to the binding energy of a single molecule. The next two terms  are obtained by taking into account  single particle-hole excitations and depend only on the atom-dimer scattering length. They are identical to the energy shift obtained for a point-like boson immersed in a Fermi sea \cite{Viverit2002gsp} and the composite nature of the dimers is only revealed by the last term of the expansion, which is calculated by taking into account two particle-hole excitations.
We stress again that, although the starting point is a variational form~\eqref{improvedansatz4}, Eq.~\eqref{ener5} is the exact low density expansion of the  molecule energy.
For comparison, let us mention that the energy obtained for a point-like boson immersed in a Fermi sea with boson-fermion scattering length  $a_{\rm ad}$
also expands as Eq.~\eqref{ener5} - from which the molecule
energy in vacuum has been subtracted - but with
$\alpha_3 \simeq  0.00025 \, \tilde{a}_{\rm ad}^3$ instead of $\alpha_3 \simeq 0.0637$ in the composite boson case.

\begin{figure}
\centerline{\includegraphics[width=\columnwidth]{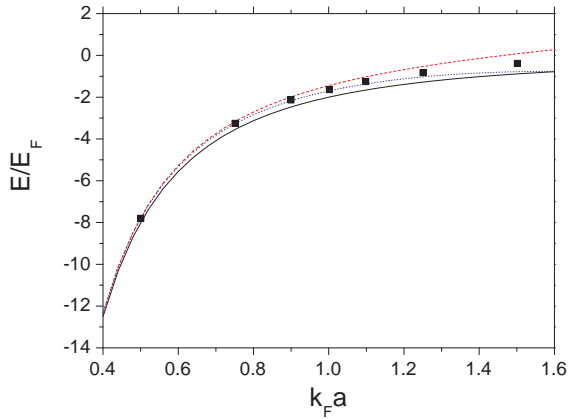}}
\caption{Perturbative expansion of the energy of a molecular impurity immersed in a Fermi sea. Square: Monte-Carlo result of \cite{prokof'ev08fpb}. Black solid line: Free molecule $E=-\hbar^2/ma^2$. Red dashed line: single particle hole excitations corresponding to the expansion up to $a^2$. Dotted blue line: expansion up to $a^3$ taking into account a second particle-hole excitation.}
\label{Fig1}
\end{figure}

\begin{figure}
\centerline{\includegraphics[width=\columnwidth]{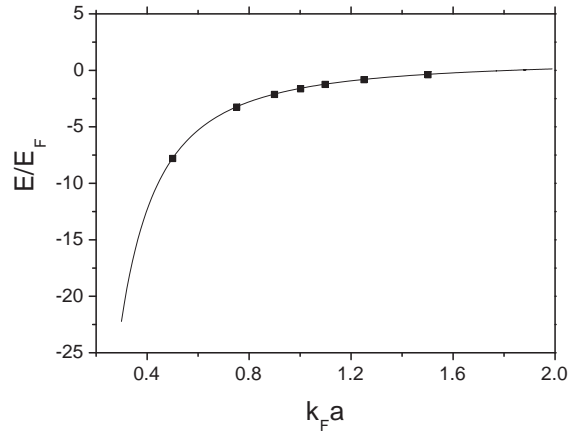}}
\caption{Variational calculation of the energy of a molecule immersed in a Fermi sea. Solid line: Variational calculation including formation of a single particle hole pair.
Points: Monte-Carlo calculation of \cite{prokof'ev08fpb}.}
\label{Fig2}
\end{figure}

It has been pointed out in \cite{prokof'ev08fpb} that the exact energy of the molecular impurity was actually very close from the mean-field correction, up to $k_F a\sim 1$. Due to its perturbative nature valid only for $k_F a\ll 1$, the analysis presented above cannot explain this intriguing feature. To address this particular issue and acquire some insight on the strongly interacting regime where the perturbative expansion diverges, we develop a variational treatment of the impurity problem. In order to get a simple and tractable calculation,
the general variational form~\eqref{improvedansatz4} is simplified by allowing only one particle-hole pair excitation and by suppressing the ${\bm q}$ dependence in
the wavefunctions. Namely, the variational state takes the form
\begin{equation}\label{improvedansatz3}
\begin{split}
& |\psi\rangle =  \Bigg( \beta  b_{{\bm K}=0}^\dagger + {\sum_{\bm k}}' A_{\bm k}
a^\dagger_{{\bm k},\downarrow} a^\dagger_{-{\bm k},\uparrow} \\[1mm]
&  - {\sum_{{\bm k},{\bm q}}}' \varphi_{{\bm k}}
a^\dagger_{{\bm k},\downarrow} b_{-{\bm k}+{\bm q}}^\dagger a_{{\bm q},\downarrow} \\[2mm]
& - \! \!   {\sum_{{\bm k},{\bm k}',{\bm q}}}'
  \Phi_{{\bm k},{\bm k'}}
a^\dagger_{{\bm k},\downarrow}
a^\dagger_{{\bm k}',\downarrow} a^\dagger_{-{\bm k'}-{\bm k}+{\bm q},\uparrow}
a_{{\bm q},\downarrow}\Bigg)|FS\rangle.
\end{split}
\end{equation}
 As shown later, this particular {\it ansatz} gives the correct expansion up to mean-field term. In addition, it can be extended to strong interactions, and the energy remains finite even at unitarity. Working out the variational equations, we reduce the problem to a set of two integral equations that are solved numerically, see Sec.~\ref{Variational}. The result is displayed in Fig. \ref{Fig2}. We observe a remarkable agreement between our simplified integral equations and the Monte-Carlo simulations. It demonstrates that, just like the Fermi-polaron problem, single particle-hole excitations provide an accurate description of the physical properties of the system.

\section{Perturbative expansion}
\label{Perturbative}

We shall detail in this Section our perturbative calculation for the immersed molecule energy.
A consistent expansion at a given order in $k_F a$ requires a corresponding minimal number of particle-hole excitations
in the variational space. We will therefore do the calculation step by step by increasing
the variational space size up to two particle-hole excitations.
The two-body problem is considered in~\ref{sec-twob}, which helps us to connect the parameters of
our starting Hamiltonian~\eqref{hamil} to the scattering length $a$ in the single-channel limit.
A trivial {\it ansatz} extending the two-body wavefunction is proposed,
which exhibits the Pauli blocking effect. Elaborating on this {\it ansatz}, single particle-hole
excitations are added in~\ref{sec-first} yielding the first (mean-field type) and second order
corrections to the molecule energy.
The third order correction requires two particle-hole excitations that are considered in~\ref{sec-second} and allows us to recover Eq. (\ref{ener5}).

\subsection{Two-body properties}\label{sec-twob}

Insight on the model~\eqref{hamil} can be gained by solving the two-body problem.
We look for a bound state of energy $E= -\frac{\hbar^2 \lambda^2}{m}$
with the {\sl ansatz}
$|\psi\rangle = ( \beta b_{{\bm K}=0}^\dagger + \sum_{\bm k} A_{\bm k}
a^\dagger_{{\bm k},\downarrow} a^\dagger_{-{\bm k},\uparrow} ) |0\rangle$.
The Schr\"odinger equation $(H-E)|\psi\rangle=0$ gives simple coupled
equations
\begin{equation}\label{2body}
(q_0^2 +\lambda^2) \beta + \sum_{\bm k} \bar{\Lambda}_k A_{\bm k} = 0 , \quad
\bar{\Lambda}_k \beta + (k^2 + \lambda^2) A_{\bm k} = 0,
\end{equation}
with $\bar{\Lambda}_k = \left( \frac{m}{\hbar^2\sqrt{V}} \right) \Lambda_k$ and
$E_0 = \frac{\hbar^2 q_0^2}{m}$. The second equation allows to express
$A_{\bm k}$ as a function of $\beta$. Replacing $A_{\bm k}$ in the first
equation yields an equation for the bound state
\begin{equation}\label{bound}
q_0^2 + \lambda^2 - \sum_{\bm k} \frac{\bar{\Lambda}_k^2}{k^2+\lambda^2} = 0.
\end{equation}
Although $\bar{\Lambda}_k$ has a very weak $k$ dependence,
the integral in Eq.~\eqref{bound} has an ultraviolet divergence if this
dependence is crudely neglected. We therefore define
$\sum_{\bm k} \frac{\bar{\Lambda}_k^2}{k^2} = \frac{V \bar{\Lambda}_0^2}{4 \pi \ell}$,
then add and substract this integral to Eq.~\eqref{bound}. $\bar{\Lambda}_k$ is
replaced by $\bar{\Lambda}_0$ for the resulting convergent integral.
$\lambda$ is finally solution of a second order equation
\begin{equation}\label{bound2}
R^* \lambda^2 + \lambda - \frac{1}{a} =0,
\end{equation}
where we define
\begin{equation}
\frac{1}{a} = \frac{1}{\ell}  - \frac{4 \pi\hbar^2 E_0}{m\Lambda_0^2 },
\quad R^* = \frac{4 \pi \hbar^4}{m^2 \Lambda_0^2}.
\end{equation}
$\ell$ acts here as a short-distance cutoff length, it is
 on the order of the van der Waals potential size.
The two-channel model Eq.~\eqref{hamil} is therefore only valid
if $R^* \gg \ell$ and $a \gg \ell$ for describing the two-body bound state.
The two-body scattering problem can also be solved using
the same {\sl ansatz}. One concludes from its solution that $a$ is
the two-body scattering length and $R^* = - r_e/2$ where $r_e$ is the
{\it effective range}.
The universal regime, or one-channel zero range model, is recovered with
$R^* \to 0$ that is $\Lambda_k \to \infty$ at fixed value of $a$.
In particular, Eq.~\eqref{bound2} gives then $\lambda = 1/a$ as expected.

It is instructive to  first  study a simple {\sl ansatz} where the dimer state
and the spin $\downarrow$ Fermi sea are not coupled, namely
\begin{equation}\label{simpleansatz}
|\psi\rangle = ( \beta b_{{\bm K}=0}^\dagger + {\sum_{\bm k}}' A_{\bm k}
a^\dagger_{{\bm k},\uparrow} a^\dagger_{-{\bm k},\downarrow} ) |FS\rangle.
\end{equation}
$|FS\rangle$ denotes the  $\downarrow$ Fermi sea with Fermi wavevector $k_F$
and density $n = k_F^3/6 \pi^2$ \footnote{This {\sl Ansatz} is actually formally identical to the textbook Cooper Ansatz used to interpret intuitively the key role of the Fermi sea in Cooper pairing in superconductors.}. The Pauli principle imposes that
wavevectors below $k_F$ are blocked  in the FS and do not participate
to the molecular state. This {\sl ansatz} is certainly not an eigenstate
of the Hamiltonian Eq.~\eqref{hamil}. The calculation that follows
is therefore variational in essence. Projecting the Schr\"odinger equation
 $(H-E)|\psi\rangle=0$ onto the two relevant sectors, two coupled
equations are derived
\begin{equation}
(q_0^2 +\lambda^2) \beta + \! \! \!
  \sum_{|\bm k|>k_F} \! \! \! \bar{\Lambda}_k A_{\bm k} = 0 , \quad
\bar{\Lambda}_k \beta + (k^2 + \lambda^2) A_{\bm k} = 0.
\end{equation}
Apart from the wavevector summation restriction, they are identical to Eqs.~\eqref{2body}.
The spin $\downarrow$ Fermi sea enters only through Pauli blocking at this stage.
The energy $E$ is measured from the Fermi sea energy.
Solving the system in the universal limit $R^* \to 0$, we find

\begin{equation}\label{lambda0}
\lambda = \frac{1}{a} - \frac{2 k_F}{\pi} \left[ 1 - \frac{\lambda}{k_F} {\rm atg}
\left( \frac{k_F}{\lambda} \right) \right].
\end{equation}

First, we can consider the dilute limit $k_F a \ll 1$,  which yields
\begin{equation}\label{enr1}
E =  - \frac{\hbar^2}{m a^2} + 2 g n + {\cal O} ( g n (k_F a)^2 ),
\end{equation}
where $g = \frac{4 \pi \hbar^2 a}{m}$. The first term in Eq.~\eqref{enr1}, {\it i.e.}
the molecule energy in vacuum, is the correct leading result in a small $k_F a$ expansion.
The second term, associated with Pauli blocking, corresponds to a mean field term treated within the Born approximation and may naively be interpreted as a first sign of the composite nature of the dimer. However, as we will see, this result is not exact,
and the correct value of the mean field term
requires the resolution of the three body problem. This can be improved by including particle-hole
excitations in the variational wavefunction as we will show next.


\subsection{First order}\label{sec-first}

Applying two times the Hamiltonian~\eqref{hamil} on the simple
previous {\sl ansatz} Eq.~\eqref{simpleansatz},
one sees that the space spanned by the variational wavefunction can be extended to
include states with a molecule and a single particle-hole excitation
from the Fermi sea. States with a closed channel boson
and a single particle-hole excitation should also be included as an intermediate step.
The improved {\sl ansatz} for the wavefunction  reads
\begin{equation}\label{improvedansatz}
\begin{split}
& |\psi\rangle =  \Bigg( \beta  b_{{\bm K}=0}^\dagger + {\sum_{\bm k}}' A_{\bm k}
a^\dagger_{{\bm k},\downarrow} a^\dagger_{-{\bm k},\uparrow} \\[1mm]
&  - {\sum_{{\bm k},{\bm q}}}' \varphi_{{\bm k},{\bm q}}
a^\dagger_{{\bm k},\downarrow} b_{-{\bm k}+{\bm q}}^\dagger a_{{\bm q},\downarrow} \\[2mm]
& - \! \!   {\sum_{{\bm k},{\bm k}',{\bm q}}}'
  \Phi_{{\bm k},{\bm k'},{\bm q}}
a^\dagger_{{\bm k},\downarrow}
a^\dagger_{{\bm k}',\downarrow} a^\dagger_{-{\bm k'}-{\bm k}+{\bm q},\uparrow}
a_{{\bm q},\downarrow} \Bigg) |FS\rangle,
\end{split}
\end{equation}
where, thanks to the linearity of Schr\"{o}dinger's equation, we can take $\beta=1$. Apart from the additional ${\bm q}$
dependence, Eq.~\eqref{improvedansatz} resembles the variational state~\eqref{improvedansatz3}.
The variational space has four different
sectors. Projecting the Schr\"odinger equation on the first two sectors, we get
\begin{subequations}\label{ftwo}
\begin{align}
(q_0^2 +\lambda^2) + \sum_{|\bm k|>k_F} \bar{\Lambda}_k A_{\bm k} &= 0 , \\[1mm]
\bar{\Lambda}_k  + (k^2 + \lambda^2) A_{\bm k}  + \sum_{|\bm q|<k_F}
\bar{\Lambda}_k\varphi_{{\bm k},{\bm q}} & = 0.
\end{align}
\end{subequations}
The aim of this calculation
is to determine how the $2 g n$ term (mean field) in Eq.~\eqref{enr1} is modified by
the improved {\sl ansatz} Eq.~\eqref{improvedansatz}.
Hence the ${\bm q}$-dependence of $\varphi_{{\bm k},{\bm q}}$
is not relevant because it leads to corrections of higher order in $k_F a$.
For $R^* \to 0$  and $k_F a \ll 1$  the modified bound state equation is found to be
\begin{equation}\label{bound3}
\lambda = \frac{1}{a} - \frac{2 k_F^3 a^2}{3 \pi} + \frac{2 k_F^3}{3 \pi} \sum_{{\bm k}}
\frac{\varphi_{\bm k}}{k^2 + 1 /a^2},
\end{equation}
where $\varphi_{\bm k}$ still  has to be determined.

Again we neglect the ${\bm q}$-dependence when projecting on the last
two sectors. Two coupled equations
are derived where the anticommuting properties of fermionic operators are essential
\begin{subequations} \label{3ch}
\begin{align}
\label{3ch1}
\begin{split}
& 2{\sum_{\bm k'}}' \bar{\Lambda}_{{\bm k}'+{\bm k}/2} \Phi_{{\bm k},{\bm k'}}
 \\[1mm]
&+ \left ( q_0^2 +\lambda^2 + \frac{3}{4} k^2  \right) \varphi_{\bm k}
= - \bar{\Lambda}_0 A_{\bm k} = \frac{\bar{\Lambda}^2_0}{k^2+\lambda^2} \\[1mm]
\end{split}
\\[2mm]
& \bar{\Lambda}_{{\bm k}'+{\bm k}/2} \left(\varphi_{\bm k}-\varphi_{\bm k'}\right) +  2\left(\lambda^2+k^2+k'^2+\bm k\cdot\bm k'\right) \Phi_{{\bm k},{\bm k'}}=0.
\end{align}
\end{subequations}

With an additional source term on the right side of Eq.~\eqref{3ch1}, this system
of equations is the same as the one derived in the three-body problem~\cite{gogolin2008asb}.
The second equation is solved readily in $\Phi_{{\bm k},{\bm k'}}$ and inserting
this result in the first equation, we obtain for $R^* \to 0$
\begin{equation}\label{inte}
\left(  \hat{L}_\lambda - \frac{1}{a} \right) \varphi_{\bm k} = \frac{4 \pi}{k^2+\lambda^2},
\end{equation}
with the usual three-body kernel~\cite{petrov2003,Skorniakov1957}
\begin{equation}
\begin{split}
\hat{L}_{\lambda} \varphi_{\bm k} \equiv &
 \sqrt{\lambda^2+3 k^2/4} \, \varphi_{\bm k} \\[1mm]
& + \frac{1}{2 \pi^2} \int d^3 k' \frac{\varphi_{{\bm k}'}}{ k^{\prime 2} +k^2+{\bm k}'\cdot {\bm k}+\lambda^2}.
\end{split}
\end{equation}
At this order, it is consistent to take $\lambda = 1/a$.
We proceed further and make contact to the atom-dimer scattering problem.
Writing $\varphi_{\bm k} = \frac{4 \pi a}{k^2} \, a_0 ( k a )$ yields
the integral equation
\begin{equation}\label{STM}
\begin{split}
& \frac{1}{\pi} \int_0^{+\infty} \, d u' \frac{a_0(u')}{u u'}
 \, \ln \left ( \frac{1+u^2+u^{\prime 2} + u u'}{1+u^2+u^{\prime 2} - u u'}
\right) \\[2mm]
& + \frac{3}{4} \frac{a_0(u)}{1+\sqrt{1+3 u^2 /4}} =
\frac{1}{1+u^2},
\end{split}
\end{equation}
identical to the one that determines the atom-dimer scattering
length~\cite{Skorniakov1957,petrov2003}.
In particular $a_0 (0) = \frac{a_{\rm ad}}{a} \simeq 1.1786\ldots$

The last term in Eq.~\eqref{bound3} is finally obtained by taking the ${\bm k}
\to {\bm 0}$ limit in Eq.~\eqref{inte},
\begin{equation}\label{correc}
\sum_{{\bm k}}
\frac{\varphi_{\bm k}}{k^2 + 1 /a^2} = a^2 \left( 1 - \frac{3}{8} \, a_0(0)
\right).
\end{equation}
This contribution is positive so that the molecule energy is lowered
by improving on the simple {\sl ansatz}. This is in fact a consequence of a general
variational principle which states that the ground state energy
can only decrease when the parameter space is extended.
Inserting Eq.~\eqref{correc} into Eq.~\eqref{bound3} leads to a final expression
for the low density molecule energy
\begin{equation}\label{enr2}
E =  - \frac{\hbar^2}{m a^2} + g_{\rm ad} n,
\end{equation}
where the atom-dimer coupling constant has been defined as
\begin{equation}\label{coupling}
 g_{\rm ad} = \frac{2 \pi \hbar^2 a_{\rm ad}}{(2 m /3)}.
\end{equation}

Interestingly, the expansion can be pushed to the next order within the same variational space.
Hence, the  ${\bm q}$ dependence of $\varphi_{{\bm k},{\bm q}}$
becomes relevant.
Skipping details, the result for $\lambda$ reads
\begin{equation}\label{lambda}
\lambda = \frac{1}{a} \left( 1 - \frac{2 x_F^3}{3 \pi}
+16 \pi^2 \sum_{|{\bm v }|<x_F} \sum_{|{\bm u}|>x_F} \frac{\tilde{\varphi}_{{\bm u},{\bm v}}}{1+u^2}
\right),
\end{equation}
where $x_F \equiv k_F a$ and $\sum_{\bm v}$ stands for $\int \frac{d^3 v}{(2 \pi)^3}$.
$\tilde{\varphi}_{{\bm u},{\bm v}}$ denotes a rescaled atom-dimer wavefunction
\begin{equation}\label{rescal}
\varphi_{{\bm k},{\bm q}} = 4 \pi a^3 \tilde{\varphi}_{{\bm k} a,{\bm q} a},
\end{equation}
solution of the integral equation
\begin{equation}\label{STM2}
\begin{split}
&  \frac{1}{2 \pi^2} \int_{|{\bm u}'|>x_F} d^3 u'
\frac{\tilde{\varphi}_{{\bm u}',{\bm v}}}{ 1+u^{\prime 2} +u^2+{\bm u}' {\bm u}
-{\bm v} ({\bm u}'+{\bm u})}  \\[1mm]
& + \left(  \sqrt{1+\frac{3 u^2}{4} -\frac{v^2}{4}-\frac{{\bm u}\cdot {\bm v}}{2}} -1 \right)
\, \tilde{\varphi}_{{\bm u},{\bm v}}
 = \frac{1}{1+u^2}
\end{split}
\end{equation}
For ${\bm v} =0$, we recover Eq.~\eqref{STM} with $\tilde{\varphi}_{{\bm u},{\bm v}} = a_0(u)/u^2$
and therefore Eq.~\eqref{enr2} for the energy.
A first idea to treat Eq.~\eqref{STM2} is to expand directly in powers of ${\bm v}$
and solve it order by order.
Note that Eq.~\eqref{lambda} involves an average over the direction of ${\bm v}$.
This would imply that the first non-vanishing correction goes as $v^2$
and therefore $x_F^5$ for the energy.
However, this approach fails because the solution of Eq.~\eqref{STM2} is singular
as ${\bm u}  \to 0$, $\tilde{\varphi}_{{\bm u},{\bm v}} \simeq a_0(0)/u^2$ for ${\bm v} =0$.

A correct treatment requires a proper description of  $\tilde{\varphi}_{{\bm u},{\bm v}}$
as a function of ${\bm v}$ for ${\bm u}  \to 0$.
We therefore write
\begin{equation}
\tilde{\varphi}_{{\bm u},{\bm v}} = \frac{a_{s1} (u,v)}{u^2 - v^2/3 - 2 {\bm u}\cdot {\bm v}/3},
\end{equation}
which translates Eq.~\eqref{STM2} to an integral equation for $a_{s1} (u,v)$.
The second term in the l.h.s. of Eq.~\eqref{STM2} can then be expanded in
${\bm v}$.
For the first term  in the l.h.s. of Eq.~\eqref{STM2}, we add and substract
the kernel for ${\bm v} =0$ and $x_F =0$. The resulting integral corrections
are simplified by changing variables, $u = v x$ and $u = x_F x$, followed
by an expansion in $v$ and $x_F$.
Keeping only first order corrections, the integral equation reads
\begin{equation}\label{firstcorr}
\begin{split}
\int_0^{+\infty} d u' K_0(u,u') a_{s1} (u',v) = \\
\frac{1+ a_{s1} (0,v)\left[ \frac{2 x_F}{\pi}
- \frac{v}{\pi} J_1(v/x_F) \right]}{1+u^2},
\end{split}
\end{equation}
with the same kernel $K_0(u,u')$ as Eq.~\eqref{STM}.
We have defined
\begin{equation}\label{js}
J_1 (s) = \frac{1}{3} \int_0^s d y  \int_{-1}^{1} d x \frac{1+ 2 x/y}{1-y^2/3-2 x y/3}.
\end{equation}
The solution of Eq.~\eqref{firstcorr} is then simply
\begin{equation}\label{solfirstcorr}
a_{s1} (u,v) = \frac{a_0 (u)}{1- a_0(0) \left( \frac{2 x_F}{\pi}
- \frac{v}{\pi} J_1(v/x_F) \right) },
\end{equation}
and can be expanded to first order in  $x_F$ and $v$.
The first outcome of this calculation is that
the singular behavior of $\tilde{\varphi}_{{\bm u},{\bm v}}$ at small
 ${\bm u}$ indeed modifies the $x_F$ (and $v$)
expansion.

The solution~\eqref{solfirstcorr} can be included in Eq.~\eqref{lambda}.
Again we add and substract the solution at vanishing $x_F$ and $v$ and
rescale integral corrections with $v$ and $x_F$.
The final result for the energy is
\begin{equation}\label{ener3}
E = - \frac{\hbar^2}{m a^2} \left( 1 - \frac{x_F^3 a_0(0)}{2 \pi}
- \frac{2 x_F^4 a_0^2(0)}{\pi^2} ( \ln 2 - 3/8) \right),
\end{equation}
which gives the next order $x_F=k_F a$ correction to Eq.~\eqref{enr2}.
We have used the following result
\begin{equation}
\int_0^1 d s \, s^3 J_1 (s) = \frac{7}{6} - \frac{4}{3} \ln 2.
\end{equation}
A few conclusion can be drawn from this result~\eqref{ener3}.
As we have mentioned already, the first correction to
the mean field result~\eqref{enr2}
is linear in $x_F$ and not quadratic.
In fact, subtracting  the dimer internal energy $-\frac{\hbar^2}{m a^2}$,
the molecule energy~\eqref{ener3} can be retrieved from
a simpler model. A point-like boson of mass $2 m$ interacting with a Fermi sea
with a boson-fermion scattering length $a_{\rm ad} = a_0(0) a$
leads also to Eq.~\eqref{ener3}.
This means that the composite structure of the boson is not apparent at
this order of the calculation.
Actually this first-order correction stems from small values of
$k \simeq k_F$. This originates from the low $k$ singular behavior in
the integral equation
and is outlined by the rescaling of variables performed during the
calculation, see above.
Hence, low $k$ corresponds to large distances over which the composite structure of
the boson is smeared out.

\subsection{Two particle-hole excitations}\label{sec-second}

The treatment becomes much more involved when pushed to next order in
$k_F$. First of all,  Pauli blocking in the two-body problem
adds a new contribution to $\lambda$. It corresponds to the $\sim k_F^5$ term
in the expansion of Eq.~\eqref{lambda0}. $\lambda$ is therefore given by
\begin{equation}\label{lambda2}
\lambda = \frac{1}{a} \left( 1 - \frac{2 x_F^3}{3 \pi}
+ \frac{2 x_F^5}{5 \pi}
+16 \pi^2 \sum_{|{\bm v }|<x_F} \sum_{|{\bm u}|>x_F} \frac{\tilde{\varphi}_{{\bm u},{\bm v}}}{1+u^2}
\right),
\end{equation}
where $\tilde{\varphi}_{{\bm u},{\bm v}}$ has to be determined.
 There are basically two kinds of contribution for the solution
of the integral equation and consequently for the energy.
The first terms come  from low $k$, {\it i.e.} values of $k$ on the
order of $k_F$ as in the last subsection. The divergence of
$\varphi_{{\bm k},{\bm q}}$ in this regime implies that one has to
go to higher orders in the calculation by including two particle-hole
excitations. Nevertheless, only the $k \to 0$ part is kept
which simplifies calculations a lot.
The second kind of terms comes from a direct expansion of
the integral equation in ${\bm v}$ once singular terms
have been carefully treated.  They involve larger values of $k$, on
the order of $1/a$.

We first need to derive a complete integral equation that includes
all terms relevant at this order of the calculation.
The wavefunction {\sl ansatz} is extended to describe two
particle-hole excitations. It is given by Eq.~\eqref{improvedansatz4} that we recall here,
\begin{equation}\label{improvedansatz2}
\begin{split}
& |\psi\rangle =  \Bigg( \beta  b_{{\bm K}=0}^\dagger + {\sum_{\bm k}}' A_{\bm k}
a^\dagger_{{\bm k},\downarrow} a^\dagger_{-{\bm k},\uparrow} \\[1mm]
&  - {\sum_{{\bm k},{\bm q}}}' \varphi_{{\bm k},{\bm q}}
a^\dagger_{{\bm k},\downarrow} b_{-{\bm k}+{\bm q}}^\dagger a_{{\bm q},\downarrow} \\[2mm]
& - \! \!   {\sum_{{\bm k},{\bm k}',{\bm q}}}'
  \Phi_{{\bm k},{\bm k'},{\bm q}}
a^\dagger_{{\bm k},\downarrow}
a^\dagger_{{\bm k}',\downarrow} a^\dagger_{-{\bm k'}-{\bm k}+{\bm q},\uparrow}
a_{{\bm q},\downarrow} \\[2mm]
& - {\sum_{{\bm k},{\bm k}',{\bm q},{\bm q}'}}' \Psi_{{\bm k},{\bm k}',{\bm q},{\bm q}'}
a^\dagger_{{\bm k},\downarrow}
a^\dagger_{{\bm k}',\downarrow} b^\dagger_{-{\bm k'}-{\bm k}+{\bm q}+{\bm q}'}
a_{{\bm q},\downarrow} a_{{\bm q}',\downarrow} \\[2mm]
& - {\sum_{{\bm k},{\bm k}',{\bm k}^{\prime \prime},{\bm q},{\bm q}'}}'
\chi_{{\bm k},{\bm k}',{\bm k}^{\prime \prime},{\bm q},{\bm q}'}
a^\dagger_{{\bm k},\downarrow}
a^\dagger_{{\bm k}',\downarrow} a^\dagger_{{\bm k}'',\downarrow} \\[2mm]
& a^\dagger_{-{\bm k'}-{\bm k}-{\bm k}^{\prime \prime}+{\bm q}+{\bm q}',\uparrow}
a_{{\bm q},\downarrow} a_{{\bm q}',\downarrow}  \Bigg) |FS\rangle,
\end{split}
\end{equation}
where all ${\bm q}$ are restricted to the Fermi surface and all ${\bm k}$
outside the Fermi surface. With no loss of generality, functions are
antisymmetrized with respect to the  ${\bm k}$ variables as well as
 with respect to the ${\bm q}$. This is of course
consistent with the anticommuting properties of the fermionic operators.

The  Schr\"odinger equation is projected on the various subspaces
of the {\sl ansatz}. Eqs.~\eqref{ftwo} are recovered while
Eqs.~\eqref{3ch} are changed to
\begin{subequations} \label{3ch2}
\begin{align}
\label{3ch12}
\begin{split}
& \left ( q_0^2 +\lambda^2 +  \tilde{\epsilon}_{\bm k}
+ \frac{\tilde{\epsilon}_{-{\bm k}+{\bm q}}}{2}
- \tilde{\epsilon}_{\bm q} \right) \varphi_{{\bm k},{\bm q}} \\[1mm]
& + 2 \sum_{|\bm k'|>k_F} \bar{\Lambda}_{{\bm k}' +{\bm k}/2} \Phi_{{\bm k},{\bm k'},{\bm q}}
= - \bar{\Lambda}_0 A_{\bm k}  ,
\end{split} \\[2mm]
\label{3ch22}
\begin{split}
&  2 \left( \tilde{\epsilon}_{\bm k} + \tilde{\epsilon}_{\bm k'}
+ \tilde{\epsilon}_{-{\bm k}-{\bm k}'+{\bm q}}
- \tilde{\epsilon}_{\bm q} + \lambda^2 \right) \Phi_{{\bm k},{\bm k'},{\bm q}} \\[1mm]
& + \bar{\Lambda}_{{\bm k}' +{\bm k}/2} ( \varphi_{{\bm k},{\bm q}} - \varphi_{{\bm k}',{\bm q}})
+ 4 \bar{\Lambda}_0 \sum_{|{\bm q}'|<k_F} \Psi_{{\bm k},{\bm k}',{\bm q},{\bm q}'}   =0,
\end{split}
\end{align}
\end{subequations}
with the notation $\tilde{\epsilon}_{\bm k} = k^2/2$. The $k$ dependence
of $\bar{\Lambda}_k$ is kept only when necessary to regularize the
two-body problem. To get an integral equation, we proceed as before.
We obtain three new terms compared to Eq.~\eqref{STM2}.
The first one is a source term proportional to
$\Psi_{{\bm k},{\bm k}',{\bm q},{\bm q}'}$.
The second comes from the $|{\bm k}'|>k_F$ restriction in Eq.~\eqref{3ch12}.
Taken for ${\bm k} \to 0$,
it gives a term proportional to $n_F \varphi_{{\bm k},{\bm q}}$.
The last term stems from the replacement of $A_{\bm k}$ using
Eqs.~\eqref{ftwo}. Taken for ${\bm k} \to 0$,
it gives a term proportional to
$\sum_{|{\bm q}'|<k_F} \varphi_{{\bm k},{\bm q}'}$.

Performing the change of variable $k = \lambda u$ followed by
the rescaling~\eqref{rescal} we obtain the integral
equation
\begin{equation}\label{STM3}
\begin{split}
&  \frac{1}{2 \pi^2} \int_{|{\bm u}'|>x_F} d^3 u'
\frac{\tilde{\varphi}_{{\bm u}',{\bm v}}}{ 1+u^{\prime 2} +u^2+{\bm u}' {\bm u}
-{\bm v} ({\bm u}'+{\bm u})}  \\[2mm]
& + \left(  \sqrt{1+\frac{3 u^2}{4} -\frac{v^2}{4}-\frac{{\bm u}\cdot {\bm v}}{2}}
- \frac{1}{a \lambda} + 4 \pi \left(\frac{x_F^3}{6 \pi^2} \right) \right)
\, \tilde{\varphi}_{{\bm u},{\bm v}}  \\[2mm]
& -  \frac{1}{2 \pi^2} \int_{|{\bm v}'|<x_F} d^3 v' \tilde{\varphi}_{{\bm u},{\bm v}'}
+ 4 \sum_{{\bm u}',{\bm v}'} \frac{\Psi_{{\bm u},{\bm u}',{\bm v},{\bm v}'}}{1+u^2}
= \frac{1}{1+u^2},
\end{split}
\end{equation}
and the energy is given by Eq.~\eqref{lambda2}.

The new corrections that we have included are all proportional to $x_F^3$
and our calculations goes up to $x_F^2$. Therefore these corrections
are negligible except for $k \simeq k_F$ (low $k$)
where they are partially compensated by
$\varphi_{{\bm k},{\bm q}} \simeq 1/k_F^2$. This justifies the
${\bm k} \to 0$ limit that is taken below.

We now project the  Schr\"odinger equation on the two remaining subspaces.
\begin{subequations} \label{3ch3}
\begin{align}
\label{3ch13}
\begin{split}
& \left ( q_0^2 +\lambda^2 +  \tilde{\epsilon}_{\bm k}  + \tilde{\epsilon}_{\bm k'}
+ \frac{\tilde{\epsilon}_{-{\bm k}-{\bm k}'+{\bm q}+{\bm q}'}}{2}
- \tilde{\epsilon}_{\bm q} - \tilde{\epsilon}_{\bm q'} \right)
\Psi_{{\bm k},{\bm k}',{\bm q},{\bm q}'}  \\[1mm]
& + 3 \sum_{{\bm k}^{\prime \prime}} \bar{\Lambda}_{{\bm k}^{\prime \prime}+{\bm k}'/2} \,
\chi_{{\bm k},{\bm k}',{\bm k}^{\prime \prime},{\bm q},{\bm q}'}
= \frac{\bar{\Lambda_0}}{2} (  \Phi_{{\bm k},{\bm k'},{\bm q}'} - \Phi_{{\bm k},{\bm k'},{\bm q}})   ,
\end{split} \\[2mm]
\label{3ch23}
\begin{split}
&  3 (\lambda^2 +  \tilde{\epsilon}_{\bm k}  +  \tilde{\epsilon}_{\bm k'}
+ \tilde{\epsilon}_{{\bm k}^{\prime \prime}}
+ \tilde{\epsilon}_{-{\bm k}-{\bm k}'-{\bm k}^{\prime \prime}+{\bm q}+{\bm q}'} \\[1mm]
& - \tilde{\epsilon}_{\bm q} - \tilde{\epsilon}_{\bm q'})
 \chi_{{\bm k},{\bm k}',{\bm k}^{\prime \prime},{\bm q},{\bm q}'}   \\[1mm]
& + \bar{\Lambda}_{{\bm k}^{\prime \prime}+{\bm k}'/2} ( \Psi_{{\bm k},{\bm k}',{\bm q},{\bm q}'}+
 \Psi_{{\bm k}',{\bm k}^{\prime \prime},{\bm q},{\bm q}'} +
 \Psi_{{\bm k}^{\prime \prime},{\bm k},,{\bm q},{\bm q}'}) =0.
\end{split}
\end{align}
\end{subequations}
These coupled equations are quite complicated. Nevertheless many simplifications
can be done that remain consistent with the order of our calculation.
The ${\bm q}$ (and  ${\bm q}'$) dependence can safely be neglected as well as
the restriction outside the Fermi sea for the ${\bm k}$ wavevectors.
The limit ${\bm k} \to 0$ is also taken.
In the source term (r.h.s.) of Eq.~\eqref{3ch13}, we can replace
$ \Phi_{{\bm k},{\bm k'},{\bm q}}$ by using its lowest order expression
from Eq.~\eqref{3ch22},
\begin{equation}
\Phi_{{\bm k},{\bm k'},{\bm q}} =  \frac{\bar{\Lambda}_0}{2}
\frac{\varphi_{{\bm k}',{\bm q}} - \varphi_{{\bm k},{\bm q}}}
{\lambda^2 + k^{\prime 2}}.
\end{equation}
Finally, Eq.~\eqref{3ch23} allows to express
 $\chi_{{\bm k},{\bm k}',{\bm k}^{\prime \prime},{\bm q},{\bm q}'}$.
Once incorporated into Eq.~\eqref{3ch13}, it leads in the
universal limit $R^* \to 0$ to
\begin{equation}
\begin{split}
& \left( \sqrt{\lambda^2 + 3 k^{\prime 2}/4} - \frac{1}{a}\right)
\Psi_{{\bm k},{\bm k}',{\bm q},{\bm q}'} \\[2mm]
& + 4 \pi \sum_{{\bm k}^{\prime \prime}}
\frac{\Psi_{{\bm k},{\bm k}^{\prime \prime},{\bm q},{\bm q}'}
+ \Psi_{{\bm k}^{\prime \prime},{\bm k}',{\bm q},{\bm q}'}}
{\lambda^2 + k^{\prime 2} + k^{\prime \prime 2} +
{\bm k}^{\prime} \cdot {\bm k}^{\prime \prime}} \\[2mm]
& = \pi \frac{\varphi_{{\bm k}',{\bm q}'} - \varphi_{{\bm k},{\bm q}'}
+ \varphi_{{\bm k},{\bm q}} - \varphi_{{\bm k}',{\bm q}}}
{\lambda^2 + k^{\prime 2}}.
\end{split}
\end{equation}
The last step is to realize that the singular behavior at low $k$,
$\varphi_{{\bm k},{\bm q}} \simeq 1/k^2$, implies that
$k$-independent terms become negligible for ${\bm k}\to0$.
Rescaling variables, {\it e.g.} ${\bm k} = \lambda u$ with $\lambda=1/a$
and Eq.~\eqref{rescal}, the following integral equation
is obtained
\begin{equation}
\begin{split}
&  4 \pi \sum_{{\bm u}^{\prime \prime}}
\frac{\Psi_{{\bm u},{\bm u}^{\prime \prime},{\bm v},{\bm v}'}}
{1 + u^{\prime 2} + u^{\prime \prime 2} +
{\bm u}^{\prime} \cdot {\bm u}^{\prime \prime}} \\[2mm]
& + \left( \sqrt{1 + 3 u^{\prime 2}/4} - 1 \right)
\Psi_{{\bm u},{\bm u}',{\bm v},{\bm v}'}
 =  \frac{4 \pi^2 ( \tilde{\varphi}_{{\bm u},{\bm v}}
- \tilde{\varphi}_{{\bm u},{\bm v}'} )}
{1 + u^{\prime 2}}.
\end{split}
\end{equation}
It acts only on ${\bm u}'$ and describes a three-body problem where $u,v,v'$ are
dummy variables.
Its solution follows trivially from the solution of Eq.~\eqref{STM},
\begin{equation}
\Psi_{{\bm u},{\bm u}',{\bm v},{\bm v}'} = 4 \pi^2 ( \tilde{\varphi}_{{\bm u},{\bm v}}
- \tilde{\varphi}_{{\bm u},{\bm v}'} ) \frac{a_0(u')}{u^{\prime 2}}.
\end{equation}
Inserting this solution into Eq.~\eqref{STM3} finally leads to the complete
integral equation
\begin{equation}\label{STM4}
\begin{split}
&  \frac{1}{2 \pi^2} \int_{|{\bm u}'|>x_F} d^3 u'
\frac{\tilde{\varphi}_{{\bm u}',{\bm v}}}{ 1+u^{\prime 2} +u^2+{\bm u}' {\bm u}
-{\bm v} ({\bm u}'+{\bm u})}  \\[2mm]
& + \left(  \sqrt{1+\frac{3 u^2}{4} -\frac{v^2}{4}-\frac{{\bm u}\cdot {\bm v}}{2}}
- 1 \right)
\, \tilde{\varphi}_{{\bm u},{\bm v}}  \\[2mm]
& -  \frac{3 a_0(0)}{16 \pi^2} \int_{|{\bm v}'|<x_F} d^3 v' \tilde{\varphi}_{{\bm u},{\bm v}'}
= \frac{1}{1+u^2}.
\end{split}
\end{equation}
Note that the $1/(a \lambda)$ term in Eq.~\eqref{STM3} has been compensated
 inside the parenthesis by using Eq.~\eqref{enr2}.

We now describe how Eq.~\eqref{STM4} is solved perturbatively.
Its solution is written in the form
\begin{equation}\label{formsol}
\begin{split}
\tilde{\varphi}_{{\bm u},{\bm v}} & = \frac{a({\bm u},{\bm v})}{u^2-v^2/3-2 {\bm u}\cdot {\bm v}/3} \\[2mm]
& \left( 1 - \frac{a_0(0)}{2 \pi^2} \int_{|{\bm v}'|<x_F} d^3 v' \frac{1}{u^2 - v^{\prime 2}/3 - 2 {\bm u}
\cdot {\bm v}'/3} \right)^{-1}
\end{split}
\end{equation}
leading to a regularized integral equation for $a({\bm u},{\bm v})$
that can be expanded in ${\bm v}$.
The integral term inside the parathesis of Eq.~\eqref{formsol} is a
correction $\sim x_F$. It is important only for small $u$ to compensate the
last term in the l.h.s. of Eq.~\eqref{STM4}.

Ordering the different source terms, $a({\bm u},{\bm v})$
can be decomposed as
\begin{equation}\label{decomp}
a({\bm u},{\bm v}) =  a_{s1} (u,v) + a_{s2} (u,v) +
a_1(u) ({\bm u}\cdot{\bm v}) + v^2 a_2(u),
\end{equation}
where the different terms are to  be detailed below.
$a_{s1} (u,v)$ is the same as in the last subsection and
is given by Eq.~\eqref{solfirstcorr}. Its lowest order in $x_F$ is
$a_0(u)$, solution of Eq.~\eqref{STM}. The first correction
in $x_F$ has been calculated  last subsection. The expansion
of Eq.~\eqref{solfirstcorr} also yields a $x_F^2$ correction.
Once plugged into Eq.~\eqref{lambda2}, $a \lambda$ receives
the correction $- a_0^3 (0) x_F^5 I_3$ with
\begin{equation}\label{I3}
I_3 =  \frac{3}{4 \pi^3} \int_{0}^{1} d s \, s^2 \, [ 2- s J_1 (s) ]^2
\simeq 0.01419,
\end{equation}
where $J_1 (s)$ is given Eq.~\eqref{js}.
The second term, $a_{s2} (u)$, corresponds to the last term in
the l.h.s. of Eq.~\eqref{STM4}. Similarly to $a_{s1} (u)$, it
stems from low $k \simeq k_F$, and reduces, up to an important
prefactor in the source term, to the three-body problem integral equation~\eqref{STM}.
The solution can be written as
\begin{equation}
 a_{s2} (u,v) = - \left( \frac{a_0(0) \, x_F}{2 \pi^2} \right)^2
a_0 (u) J_2 \left( \frac{v}{x_F} \right),
\end{equation}
where
\begin{equation}
J_2 (y) = \int_{x >1} \! \! \! \! d^3 x \frac{1}{x^2- \frac{y^2}{3}
- \frac{2 {\bm x} \cdot {\bm y}}{3}}
\int_{y' <1} \! \! \! \! d^3 y' \frac{1}{x^2 - \frac{y^{\prime 2}}{3}
-   \frac{2 {\bm x} \cdot {\bm y}'}{3}}.
\end{equation}
The contribution to $a \lambda$ is given by $a_0^3 (0) x_F^5 I_2$
with the integral
\begin{equation}\label{I2}
I_2  = \frac{3}{16 \pi^5} \int_0^1 d y \, y^2 J_2 (y) \simeq 0.014440.
\end{equation}
In a way similar to the first order correction,
the two contributions stemming from $a_{s1}$ and $a_{s2}$
are produced by small wavevectors $k \simeq k_F$ and
match exactly the total $x_F^2$ correction of the point-like boson model.
They correspond to the large distance part for the scattering
of the boson by the Fermi sea. On this scale $k \simeq k_F$,
the composite structure of the boson is not apparent.

This is in contrast with the situation for
the last two terms in Eq.~\eqref{decomp}. They result from the
direct expansion in $v$ of the integral equation.
In that case, typical wavevectors $k$ are of order $1/a \gg k_F$.
These terms involve the whole spatial structure of the composite molecule.
They have no equivalent in the point-like boson model and hence
reveal the composite nature of the molecule.
The integral equations that determine $a_1(u)$ and $a_2(u)$ follow
from a tedious but systematic expansion in $v$.
\begin{equation}
\int_0^{+\infty} d u' K_1(u,u') a_1(u') = S_1(u)
\end{equation}
determines $a_1(u)$ with the Kernel
\begin{equation}
K_1(u,u') = \frac{3}{4} \frac{\delta(u-u')}{1+E}
+ \frac{1}{\pi} \frac{u^{\prime 2} }{B^2} \left[2 - \frac{A}{B}
\ln \left( \frac{A+B}{A-B} \right) \right],
\end{equation}
with $A = 1+u^2 + u^{\prime 2}$, $B= u u'$ and $E = \sqrt{1+3 u^2/4}$.
Using the same definitions, the source term can be written
\begin{equation}
\begin{split}
&  S_1(u) = - \frac{3}{16} \frac{ a_0(u)}{E (1+E)^2} \\[1mm]
&  - \frac{2}{\pi} \int_0^{+\infty} d u' \frac{a_0(u')}{u^2}
\left[ \frac{u^2-A}{A^2-B^2} + \frac{1}{2B}
\ln  \left(\frac{A+B}{A-B} \right) \right] \\[1mm]
& - \frac{2}{3 \pi} \int_0^{+\infty} d u' \frac{a_0(u')}{B^2}
\left[2 - \frac{A}{B}
\ln \left( \frac{A+B}{A-B} \right) \right].
\end{split}
\end{equation}
The integral equation for $a_2(u)$,
\begin{equation}
\int_0^{+\infty} d u' K_0(u,u') a_2(u') = S_2^{(1)}(u) + S_2^{(2)} (u),
\end{equation}
has the same Kernel as Eq.~\eqref{STM} with two sources terms.
Note that this result has already been averaged over the directions
of ${\bm v}$ and ${\bm u}$ since only that component matters
for $\lambda$, see Eq.~\eqref{lambda2}.

The first source term, $S_2^{(1)}(u)$, depends on $a_1(u)$
\begin{equation}
\begin{split}
& S_2^{(1)}(u)  = - \frac{1}{16} \frac{u^2 a_1(u)}{E (1+E)^2}  \\[1mm]
& - \frac{2}{3 \pi} \int_0^{+\infty} d u' a_1(u')
\left[ \frac{u^{\prime 2}-A}{A^2-B^2} + \frac{1}{2B}
\ln  \left(\frac{A+B}{A-B} \right) \right] \\[1mm]
& - \frac{2}{9 \pi} \int_0^{+\infty} d u' a_1(u')
\frac{1}{B} \ln  \left(\frac{A+B}{A-B} \right).
\end{split}
\end{equation}
The second source term depends on $a_0(u)$,
\begin{equation}
\begin{split}
& S_2^{(2)}(u)  =  - \frac{3 a_0(u)}{32 E (1+E)^2}
- \frac{u^2 (1+3 E) a_0(u)}{128 E^3 (1+E)^3} \\[1mm]
& - \frac{26}{27 \pi} \int_0^{+\infty} \frac{d u'}{u^{\prime 2}}
\left[\frac{a_0(u')}{2 B} \ln  \left(\frac{A+B}{A-B} \right)
- \frac{a_0(0)}{1+u^2} \right] \\[1mm]
& - \frac{4}{9 \pi} \int_0^{+\infty} d u' \frac{a_0(u')}{u^{\prime 2}}
\left[ \frac{u^{\prime 2}-A}{A^2-B^2} + \frac{1}{2B}
\ln  \left(\frac{A+B}{A-B} \right) \right]\\[1mm]
& - \frac{2}{3 \pi} \int_0^{+\infty} d u' a_0(u')
\frac{(u^2+u^{\prime 2})A -B^2}{(A^2-B^2)^2}.
\end{split}
\end{equation}
The resulting contribution to $a \lambda$, Eq.~\eqref{lambda2}, is
added to the Pauli blocking term (third term inside the parenthesis
of Eq.~\eqref{lambda2}) to give $x_F^5 I_1$ with
\begin{equation}\label{I1}
\begin{split}
& I_1 = \frac{2}{5 \pi} -\frac{3 a_2(0)}{20 \pi } - \frac{3  a_0(0)}{320 \pi}
- \frac{4}{15 \pi^2}\int_0^{+\infty} du \frac{u^2 a_1(u)}{(1+u^2)^2} \\[2mm]
& - \frac{8}{45 \pi^2} \int_0^{+\infty} du \frac{a_0(u) (1+5 u^2/2)}{(1+u^2)^2}
 \simeq 0.063324.
\end{split}
\end{equation}

Gathering all $x_F^2$ corrections to the mean field result, we finally
obtain that Eq.~\eqref{ener3} is extended to
\begin{equation}\label{ener4}
\begin{split}
E  = - \frac{\hbar^2}{m a^2} & \Bigg( 1 - \frac{x_F^3 a_0(0)}{2 \pi}
- \frac{2 x_F^4 a_0^2(0)}{\pi^2} ( \ln 2 - 3/8) \\[2mm]
& + 2 x_F^5 [ a_0^3 (0) (I_2 - I_3) + I_1 ] \Bigg),
\end{split}
\end{equation}
with Eqs.~\eqref{I1},\eqref{I2},\eqref{I3} for $I_1, I_2$ and $I_3$.
Note that $a_0^3 (0) (I_2 - I_3) + I_1 \simeq 0.0637$.
It indicates that the contribution $I_1$, specific to the composite
boson, dominates quantitatively the last term in the energy.
The point-like boson model also leads to Eq.~\eqref{ener4} but with $I_1=0$.
The final result~\eqref{ener4} was announced in Sec.~\ref{Main} as Eq.~\eqref{ener5}.

\section{Variational Treatment}
\label{Variational}

As mentioned in Sec.~\ref{Main} where the main results of this article are summarized, an alternative way of treating the problem is to use a variational approach in the restricted subspace given by Eq.~\eqref{improvedansatz3}. We recall  here the variational form
\begin{equation}\label{improvedansatz3b}
\begin{split}
& |\psi\rangle =  \Big(\beta  b_{{\bm K}=0}^\dagger + {\sum_{\bm k}}' A_{\bm k}
a^\dagger_{{\bm k},\downarrow} a^\dagger_{-{\bm k},\uparrow} \\[1mm]
&  - {\sum_{{\bm k},{\bm q}}}' \varphi_{{\bm k}}
a^\dagger_{{\bm k},\downarrow} b_{-{\bm k}+{\bm q}}^\dagger a_{{\bm q},\downarrow} \\[2mm]
& - \! \!   {\sum_{{\bm k},{\bm k}',{\bm q}}}'
  \Phi_{{\bm k},{\bm k'}}
a^\dagger_{{\bm k},\downarrow}
a^\dagger_{{\bm k}',\downarrow} a^\dagger_{-{\bm k'}-{\bm k}+{\bm q},\uparrow}
a_{{\bm q},\downarrow}\Big)|FS\rangle,
\end{split}
\end{equation}
where $\Phi_{{\bm k},{\bm k'}}$ has been antisymmetrized.
As observed in the previous section~\ref{Perturbative},  this {\it ansatz} is sufficient to recover the mean-field correction~\eqref{enr2} corresponding to the first order term in the perturbative expansion~\eqref{ener4}. Minimizing the energy with respect to the amplitudes $\beta$, $A_k$, $\varphi_k$ and $\Phi_{\bm k,\bm k'}$, while keeping $\langle \psi|\psi\rangle$ constant leads to a set of equations very similar to the one obtained in the perturbative treatment of the problem

\begin{eqnarray}
(E_0-E)\beta +\frac{1}{\sqrt{V}}{\sum_{\bm k}}'\Lambda_kA_k=0\label{VariationalEqn1}\\
(2\varepsilon_k-E)A_k+\frac{\Lambda_k}{\sqrt{V}}\beta+\frac{1}{\sqrt{V}}{\sum_{\bm q}}'\Lambda_k\varphi_k=0\label{VariationalEqn2}\\
\begin{split}
{\sum_{\bm q}}'(E_0+\varepsilon_{\bm k}+\varepsilon_{\bm q-\bm k}/2-\varepsilon_{\bm q}-E)\varphi_k+\\
{\sum_{\bm q}}'\frac{\Lambda_k}{\sqrt{V}}A_k+\frac{2}{\sqrt{V}}{\sum_{\bm k',\bm q}}'\Lambda_{{\bm k}'+{\bm k}/2} \Phi_{\bm k,\bm k'} =0\label{VariationalEqn3}
\end{split}
\\
\begin{split}
{\sum_{\bm q'}}'(\varepsilon_k+\varepsilon_{k'}+\varepsilon_{\bm q-\bm k-\bm k'}-\varepsilon_{\bm q}-E)\Phi_{\bm k,\bm k'}+\\
\frac{1}{2\sqrt{V}}{\sum_{\bm q}}' \left( \Lambda_{{\bm k}'+{\bm k}/2} \, \varphi_{\bm k}-
\Lambda_{{\bm k}+{\bm k}'/2} \, \varphi_{\bm k'}\right)=0 \label{VariationalEqn4}
\end{split}
\end{eqnarray}

Note that here the energy $E$ appears as the Lagrange multiplier associated with the constraint on $\langle \psi|\psi\rangle$. Moreover, since the equations are linear, we can set $\beta=1$.

The first two equations are identical to those we used in the perturbative calculation. Introducing $\lambda$ defined by $E=-\hbar^2\lambda^2/m$, it yields the first variational equation:

\be
\lambda = \frac{1}{a}-\frac{2k_F}{\pi}\left[1-\frac{\lambda}{k_F}\arctan (k_F/\lambda)\right]+\frac{2k_F^3}{3\pi}{\sum_{\bm k}}'\frac{\varphi_k}{k^2+\lambda^2}.
\label{EqnVariational1}
\ee

In Eq. (\ref{VariationalEqn4}), the explicit $q$ dependence vanishes with the radial sum, and we finally recover the equation

\be
\Phi_{\bm k,\bm k'}=\frac{1}{2\sqrt{V}}\left( \frac{\Lambda_{{\bm k}+{\bm k}'/2}\,\varphi_{\bm k'}-\Lambda_{{\bm k}'+{\bm k}/2} \, \varphi_{\bm k}}{\varepsilon_{\bm k}+\varepsilon_{\bm k'}+\varepsilon_{\bm k+\bm k'}-E}\right).
\label{EqnPhi}
\ee

Inserting Eq. \eqref{VariationalEqn2} and \eqref{EqnPhi} in \eqref{VariationalEqn3} finally yields in the short range limit a closed equation for the field $\varphi$:

\be
\left(-\frac{1}{4\pi a}+F_k+\widehat L'_k\right)\varphi_k=\frac{1}{V} \frac{1}{2\varepsilon_k-E},
\label{EqnVariational2}
\ee

\noindent where the function $F$ and the integral kernel $\widehat L'$ are defined by

\be
\widehat L'\varphi_k=
\frac{1}{V}{\sum_{\bm k'}}'\frac{\varphi_{\bm k'}}{\varepsilon_{\bm k}+\varepsilon_{\bm k'}+\varepsilon_{\bm k+\bm k'}-E}\\
\ee

\be
\begin{split}
F_k=& \frac{1}{V}{\sum_{\bm q}}'\frac{1}{E-2\varepsilon_k}+\frac{1}{4 \pi} \sqrt{\lambda^2+ \frac{3}{4} k^2}\\
&+\frac{1}{V} \sum_{|{\bm k}'|<k_F} \left(\frac{1}{\varepsilon_{\bm k}+\varepsilon_{\bm k'}+\varepsilon_{\bm k+\bm k'}-E}\right)
\end{split}
\ee

Eq. (\ref{EqnVariational1}) and (\ref{EqnVariational2}) are then solved numerically, yielding the result presented in Fig. (\ref{Fig2}).
 A  good agreement between our variational {\it ansatz} and the exact Monte-Carlo simulations is obtained. As already noted, it is very close to the mean-field prediction up to $k_Fa\sim 1$. However, it stays finite at unitarity, since we have  $E\simeq 1.1506 \, E_F$ for $k_F a=\infty$.

\section{Conclusion}
\label{conclusion}

In analogy with the Fermi-polaron system, the work presented here shows that the molecular sector of the impurity problem can be described quantitatively as a molecule dressed by a single particle-hole excitation. However, two important points are still to be clarified and will be addressed in future work. First, what is the effective mass of the quasi-particle ? This property is in particular important to capture the dynamical behavior of the system, as suggested \cite{lobo2006nsp} and observed recently in experiments \cite{Nascimbene2009}. Moreover, the bosonic/fermionic nature of the quasi-particle should be clarified by the study of an ensemble of impurities immersed in a Fermi sea, that may help clarifying  the molecule-polaron transition scenario. This formalism can also be used to interpret spectroscopic data obtained for instance in Ref.~\cite{schirotzek2009ofp} and refine the Nozi\`eres Schmitt-Rink analysis of Ref.~\cite{massignan08spm}. Another extension of this work deals with mixtures of particles with unequal masses, for instance the Li-K mixture.

We thank N. Navon, S. Nascimb\`ene, C. Lobo and P. Massignan for fruitful discussions.
F.C. acknowledges support from ESF (SCALA and EuroQUAM),
ANR FABIOLA and GasCor, R\'egion Ile de France (IFRAF), ERC
Ferlodim and Institut Universitaire de France.

\bibliography{Imbalanced}

\end{document}